# On separable Schrödinger equations

Renat Zhdanov[*] and Alexander Zhalij[†]

Institute of Mathematics of the Academy of Sciences of Ukraine,

Tereshchenkivska Street 3, 252004 Kyiv, Ukraine

## Abstract

We classify (1+3)-dimensional Schrödinger equations for a particle interacting with the electromagnetic field that are solvable by the method of separation of variables. As a result, we get eleven classes of the electromagnetic vector potentials of the electromagnetic field $A(t,\vec{x}) = (A_0(t,\vec{x}), \vec{A}(t,\vec{x}))$ providing separability of the corresponding Schrödinger equations. It is established, in particular, that the necessary condition for the Schrödinger equation to be separable is that the magnetic field must be independent of the spatial variables. Next, we prove that any Schrödinger equation admitting variable separation into second-order ordinary differential equations can be reduced to one of the eleven separable Schrödinger equations mentioned above and carry out variable separation in the latter. Furthermore, we apply the results obtained for separating variables in the Hamilton–Jacobi equation.

## I. Introduction

Being invented by Fourier and Euler long ago, the method of separation of variables is still the most powerful and efficient one for integrating linear partial differential equations (PDEs). This is especially the case for PDEs having variable coefficients, where the standard Fourier transform is no longer applicable. Moreover, this method proves to be a useful tool for constructing particular solutions of some nonlinear partial differential equations such as the nonlinear Laplace [1], wave [2, 3], and heat conductivity equations [4]–[6] in (1+1) dimensions.

---

[*]e-mail: renat@imath.kiev.ua

[†]e-mail: zhaliy@imath.kiev.ua

The principal object of study in the present paper is a problem of the separation of variables in the Schrödinger equation (SE) for a particle interacting with the electromagnetic field,

$$(p_0 - p_a p_a)\psi(t, \vec{x}) = 0. \tag{1}$$

Here we use the notations

$$p_0 = i\frac{\partial}{\partial t} - eA_0(t, \vec{x}), \quad p_a = i\frac{\partial}{\partial x_a} - eA_a(t, \vec{x}), \quad a = 1, 2, 3,$$

where $A = (A_0, A_1, A_2, A_3)$ is the vector potential of the electromagnetic field, $e = $ const. Hereafter the summation over the repeated indices from 1 to 3 is understood.

Böcher is believed to be the first to obtain in [7] a systematic classification of coordinate systems enabling separability of the three-dimensional stationary SE,

$$(-\Delta + E)\psi(\vec{x}) = 0, \tag{2}$$

where $\Delta$ is the Laplacian in three dimensions.

He has shown that equation (2) is separable via the separation Ansatz,

$$\psi(\vec{x}) = \varphi_1(\omega_1(\vec{x}))\varphi_2(\omega_2(\vec{x}))\varphi_3(\omega_3(\vec{x})) \tag{3}$$

in eleven inequivalent coordinate systems $\omega_1(\vec{x})$, $\omega_2(\vec{x})$, $\omega_3(\vec{x})$. We list these following the famous paper by Eisenhart [8], where a rigorous geometric derivation of the corresponding results is given. Note that the coordinate systems are given in the implicit form $x_a = z_a(\omega_1, \omega_2, \omega_3)$, $a = 1, 2, 3$.

1. Cartesian coordinate system,

   $z_1 = \omega_1, \quad z_2 = \omega_2, \quad z_3 = \omega_3,$

   $\omega_1, \omega_2, \omega_3 \in \mathbf{R}.$

2. Cylindrical coordinate system,

$$z_1 = e^{\omega_1}\cos\omega_2, \quad z_2 = e^{\omega_1}\sin\omega_2, \quad z_3 = \omega_3,$$
$$0 \leq \omega_2 < 2\pi, \quad \omega_1, \omega_3 \in \mathbf{R}.$$

3. Parabolic cylindrical coordinate system,
$$z_1 = (\omega_1^2 - \omega_2^2)/2, \quad z_2 = \omega_1\omega_2, \quad z_3 = \omega_3,$$
$$\omega_1 > 0, \quad \omega_2, \omega_3 \in \mathbf{R}.$$

4. Elliptic cylindrical coordinate system,
$$z_1 = a\cosh\omega_1\cos\omega_2, \quad z_2 = a\sinh\omega_1\sin\omega_2, \quad z_3 = \omega_3,$$
$$\omega_1 > 0, \quad -\pi < \omega_2 \leq \pi, \quad \omega_3 \in \mathbf{R}, \quad a > 0.$$

5. Spherical coordinate system,
$$z_1 = \omega_1^{-1}\operatorname{sech}\omega_2\cos\omega_3,$$
$$z_2 = \omega_1^{-1}\operatorname{sech}\omega_2\sin\omega_3,$$
$$z_3 = \omega_1^{-1}\tanh\omega_2,$$
$$\omega_1 > 0, \quad \omega_2 \in \mathbf{R}, \quad 0 \leq \omega_3 < 2\pi.$$

6. Prolate spheroidal coordinate system,
$$z_1 = a\operatorname{csch}\omega_1\operatorname{sech}\omega_2\cos\omega_3, \quad a > 0,$$
$$z_2 = a\operatorname{csch}\omega_1\operatorname{sech}\omega_2\sin\omega_3,$$
$$z_3 = a\coth\omega_1\tanh\omega_2, \tag{4}$$
$$\omega_1 > 0, \quad \omega_2 \in \mathbf{R}, \quad 0 \leq \omega_3 < 2\pi.$$

7. Oblate spheroidal coordinate system,
$$z_1 = a\csc\omega_1\operatorname{sech}\omega_2\cos\omega_3, \quad a > 0,$$
$$z_2 = a\csc\omega_1\operatorname{sech}\omega_2\sin\omega_3,$$
$$z_3 = a\cot\omega_1\tanh\omega_2,$$
$$0 < \omega_1 < \pi/2, \quad \omega_2 \in \mathbf{R}, \quad 0 \leq \omega_3 < 2\pi.$$

8. Parabolic coordinate system,
$$z_1 = e^{\omega_1+\omega_2}\cos\omega_3, \quad z_2 = e^{\omega_1+\omega_2}\sin\omega_3,$$
$$z_3 = (e^{2\omega_1} - e^{2\omega_2})/2,$$
$$\omega_1, \omega_2 \in \mathbf{R}, \quad 0 \leq \omega_3 \leq 2\pi.$$

9. Paraboloidal coordinate system,
$$z_1 = 2a\cosh\omega_1\cos\omega_2\sinh\omega_3, \quad a > 0,$$
$$z_2 = 2a\sinh\omega_1\sin\omega_2\cosh\omega_3,$$

$$z_3 = a(\cosh 2\omega_1 + \cos 2\omega_2 - \cosh 2\omega_3)/2,$$
$$\omega_1, \omega_3 \in \mathbf{R}, \quad 0 \leq \omega_2 < \pi.$$

10. Ellipsoidal coordinate system,
$$z_1 = a\,\frac{1}{\operatorname{sn}(\omega_1,\,k)}\,\operatorname{dn}(\omega_2,\,k')\operatorname{sn}(\omega_3,\,k), \quad a > 0,$$
$$z_2 = a\,\frac{\operatorname{dn}(\omega_1,\,k)}{\operatorname{sn}(\omega_1,\,k)}\,\operatorname{cn}(\omega_2,\,k')\operatorname{cn}(\omega_3,\,k),$$
$$z_3 = a\,\frac{\operatorname{cn}(\omega_1,\,k)}{\operatorname{sn}(\omega_1,\,k)}\,\operatorname{sn}(\omega_2,\,k')\operatorname{dn}(\omega_3,\,k),$$
$$0 < \omega_1 < K, \quad -K' \leq \omega_2 \leq K', \quad 0 \leq \omega_3 \leq 4K.$$

11. Conical coordinate system,
$$z_1 = \omega_1^{-1}\operatorname{dn}(\omega_2,\,k')\operatorname{sn}(\omega_3,\,k),$$
$$z_2 = \omega_1^{-1}\operatorname{cn}(\omega_2,\,k')\operatorname{cn}(\omega_3,\,k),$$
$$z_3 = \omega_1^{-1}\operatorname{sn}(\omega_2,\,k')\operatorname{dn}(\omega_3,\,k),$$
$$\omega_1 > 0, \quad -K' \leq \omega_2 \leq K', \quad 0 \leq \omega_3 \leq 4K.$$

Here we use the usual notations for the trigonometric, hyperbolic and Jacobi elliptic functions, $k\,(0 < k < 1)$ being the modulus of the latter and $k' = (1 - k^2)^{1/2}$.

By the evident reasons, the coordinate systems 1, 2–4, and 5–11 from the above list are called completely split, partially split, and non–split, correspondingly.

Note that the above list differs slightly from the one presented in [8], since we have rearranged the coordinate systems $\omega_1, \omega_2, \omega_3$ in such a way that the relations
$$\Delta\omega_a = 0, \quad a = 1, 2, 3 \tag{5}$$
hold for all the cases 1–11.

Next, for each of the coordinate systems, Eisenhart [9] determined the form of the potential $V(\vec{x})$ that permits the separation of variables.

The Eisenhart's technique has been applied by Olevskii [10] for separating variables in the Laplace–Beltrami operator in the spaces of constant curvature (see, also [11]).

Smorodinsky and Winternitz with co-workers [12, 13] started a systematic study of potentials for which the stationary SE in two and three dimensions admits the separation of variables in two or more coordinate systems (so-called superintegrable potentials). The classification of these potentials has been completed by Evans [14].

In the mid-seventies a series of papers by Miller and Kalnins appears, where a symmetry approach to variable separation has been developed. This approach is based on the well-known fact that a solution with separated variables is a common eigenfunction of first- or second-order differential operators, which commute with each other and with the operator of an equation under consideration. Further details and an extensive list of references can be found in the monograph [15] (also see, the review by Koornwinder [16]). Boyer, Kalnins, and Miller have obtained a systematic solution of the variable separation problem in the time-dependent (1+2)-dimensional free SE within the framework of the symmetry approach [17]. Later on, Boyer has described all coordinate systems providing separability of the (1+2)-dimensional SE having the potential $V(x_1, x_2) = \alpha/x_1^2 + \beta/x_2^2$ [18]. Reid has completely solved the problem of variable separation in the three-dimensional time-dependent SE for a free particle [19].

Independently, the symmetry approach to the separation of variables in the equations of quantum mechanics and quantum field theory was developed by Shapovalov, who was the first to give a systematic treatment of the problem of variable separation in the Dirac equation using its non-Lie symmetry [20] and by Bagrov with collaborators (see, [21] and references therein). Shapovalov and Sukhomlin [22] have obtained some separable SEs of the form (1), however, their results are not complete. Let us also mention the papers [23, 24], where the physical aspects of the problem of separation of variables in some (1+3)- and (1+1)-dimensional SE with time dependent potentials are studied.

In the present paper we are mainly devoted to the inverse problem of variable separation in SE (1), namely, to one of the classifying PDEs of the form (1) that can be solved by the method of separation of variables. Clearly, to be able to handle this problem efficiently, we need a precise algorithmic definition of what the separation of variables is. We have suggested a possible definition of the separation of variables applicable both to linear [25, 26] and nonlinear [27] PDEs, which enables developing an efficient approach to solving classification problems for (1+2)-dimensional time-dependent Schrödinger equations having time-independent [26, 27] and time-dependent scalar potentials [28]. Recently, we have obtained an exhaus-tive classification of separable Schrödinger equations (1) in (1+2) dimensions with $A_3 = 0$ [29].

In the present paper we solve completely the problem of variable separation in SE (1) into second-order ordinary differential equations in a sense that we obtain all possible forms of the vector potential $A(t, \vec{x}) = (A_0(t, \vec{x}), A_1(t, \vec{x}), A_2(t, \vec{x}), A_3(t, \vec{x}))$ providing separability of (1). Furthermore, we construct inequivalent coordinate systems enabling us to separate variables in the corresponding SEs and carry out variable separation.

The paper has the following structure. In the second section we solve the classification problem for SE (1) and obtain all vector functions $A(t, \vec{x})$ such that (1) is separable. What it is more, we consider briefly the problem of the separation of variables in SE having the generalized Coulomb potential and construct all the possible coordinate systems providing its separability. The third Section is devoted to the application of the results of Section II to separate variables in the (1+3)-dimensional Hamilton–Jacobi equation. In a concluding section we indicate some further applications of the results obtained in the paper.

## II. Separation of variables in the Schrödinger equation

With all the variety of approaches to a separation of variables in PDEs, one can notice the three generic principles respected by all of them, namely, the following

1. Representation of a solution to be found in a separated (factorized) form via several functions of one variable.

2. Requirement that the above mentioned functions of one variable should satisfy some ordinary differential equations.

3. Dependence of the so found solution on several arbitrary (continuous or discrete) parameters, called spectral parameters or separation constants.

By a proper formalizing the above features we have formulated in [27] an algorithm for variable separation in linear PDEs. Below we apply this algorithm in order to classify separable SEs of the form (1).

We say that SE (1) is separable in a coordinate system $t$, $\omega_a = \omega_a(t, \vec{x})$, $a = 1, 2, 3$ if the separation Ansätz,

$$\psi(t, \vec{x}) = Q(t, \vec{x})\varphi_0(t) \prod_{a=1}^{3} \varphi_a\left(\omega_a(t, \vec{x}), \vec{\lambda}\right), \qquad (6)$$

reduces PDE (1) to four ordinary differential equations for the functions $\varphi_\mu$, $(\mu = 0, 1, 2, 3)$,

$$\varphi_0' = U_0(t, \varphi_0; \vec{\lambda}), \quad \varphi_a'' = U_a(\omega_a, \varphi_a, \varphi_a'; \vec{\lambda}). \qquad (7)$$

Here $U_0, \ldots, U_3$ are some smooth functions of the indicated variables, $\vec{\lambda} = (\lambda_1, \lambda_2, \lambda_3) \in \Lambda = \{$an open domain in $\mathbf{R}^3\}$ are separation constants and, what is more,

$$\operatorname{rank} \left\| \begin{array}{ccc} \frac{\partial U_0}{\partial \lambda_1} & \frac{\partial U_0}{\partial \lambda_2} & \frac{\partial U_0}{\partial \lambda_3} \\ \frac{\partial U_1}{\partial \lambda_1} & \frac{\partial U_1}{\partial \lambda_2} & \frac{\partial U_1}{\partial \lambda_3} \\ \frac{\partial U_2}{\partial \lambda_1} & \frac{\partial U_2}{\partial \lambda_2} & \frac{\partial U_2}{\partial \lambda_3} \\ \frac{\partial U_3}{\partial \lambda_1} & \frac{\partial U_3}{\partial \lambda_2} & \frac{\partial U_3}{\partial \lambda_3} \end{array} \right\| = 3. \qquad (8)$$

The above condition secures the essential dependence of a solution with separated variables on the separation constants $\vec{\lambda}$.

Formulas (6)–(8) form the input data of the method. The principal steps of the procedure of variable separation in SE (1) are as follows.

1. We insert the Ansätz (6) into SE and express the derivatives $\varphi_0'$, $\varphi_1''$, $\varphi_2''$, $\varphi_3''$ in terms of functions $\varphi_0$, $\varphi_1$, $\varphi_2$, $\varphi_3$, $\varphi_1'$, $\varphi_2'$, $\varphi_3'$ using equations (7).

2. We regard $\varphi_0$, $\varphi_1$, $\varphi_2$, $\varphi_3$, $\varphi_1'$, $\varphi_2'$, $\varphi_3'$, $\lambda_1$, $\lambda_2$, $\lambda_3$ as the new independent variables $y_1, \ldots, y_{10}$. As the functions $Q$, $\omega_1$, $\omega_2$, $\omega_3$ are independent of the variables $y_1, \ldots, y_{10}$ we can split by these variables and get an overdetermined system of nonlinear partial differential equations for unknown functions $Q$, $\omega_1$, $\omega_2$, $\omega_3$.

3. After solving the above system we get an exhaustive description of coordinate systems providing separability of SE.

Having performed the first two steps of the above algorithm, we arrive at the conclusion that the separation equations (7) are linear both in $\varphi_0, \ldots, \varphi_3$ and $\lambda_1, \lambda_2, \lambda_3$ (the principal reason for this is the fact that SE (1) is linear).

Next, we introduce an equivalence relation $\mathcal{E}$ on the set of all coordinate systems providing the separability of SE. We say that two coordinate systems $t, \omega_1, \omega_2, \omega_3$ and $\tilde{t}, \tilde{\omega}_1, \tilde{\omega}_2, \tilde{\omega}_3$ are equivalent if the corresponding Ansätzes (6) are transformed one into another by

- the group transformations from the Lie transformation group admitted by SE (1) with $A = 0$,
- the invertible transformations of the form

$$t \to \tilde{t} = f_0(t), \quad \omega_i \to \tilde{\omega}_i = f_i(\omega_i), \tag{9}$$
$$Q \to \tilde{Q} = Q l_0(t) l_1(\omega_1) l_2(\omega_2) l_3(\omega_3), \tag{10}$$

where $f_0, \ldots, f_3, l_0, \ldots, l_3$ are some smooth functions and $i = 1, 2, 3$.

This equivalence relation reflects the freedom in the choice of the functions $Q, \omega_1, \omega_2, \omega_3$ and separation constants $\lambda_1, \lambda_2, \lambda_3$ preserving the form of the separation Ansatz (6). It splits the set of all possible coordinate systems into equivalence classes. In a sequel, when presenting the lists of coordinate systems enabling us to separate variables in SE we will give only one representative for each equivalence class.

Within the equivalence relation $\mathcal{E}$ we can always choose the reduced equations (7) to be

$$i\varphi_0' = (T_0(t) - T_i(t)\lambda_i)\varphi_0, \quad \varphi_a'' = (F_{a0}(\omega_a) + F_{ai}(\omega_a)\lambda_i)\varphi_a, \tag{11}$$

where $T_0, T_i, F_{a0}, F_{ai}$ are some smooth functions of the indicated variables, $a = 1, 2, 3$. With this remark the system of nonlinear PDEs for unknown functions $Q, \omega_1, \omega_2, \omega_3$ takes the form

$$\frac{\partial \omega_i}{\partial x_a} \frac{\partial \omega_j}{\partial x_a} = 0, \quad i \neq j, \quad i, j = 1, 2, 3; \tag{12}$$

$$\sum_{i=1}^{3} F_{ia}(\omega_i) \frac{\partial \omega_i}{\partial x_j} \frac{\partial \omega_i}{\partial x_j} = T_a(t), \quad a = 1, 2, 3; \tag{13}$$

$$2\left(\frac{\partial Q}{\partial x_j} + ieQA_j\right)\frac{\partial \omega_a}{\partial x_j} + Q\left(i\frac{\partial \omega_a}{\partial t} + \Delta\omega_a\right) = 0, \quad a = 1, 2, 3; \tag{14}$$

$$Q\sum_{i=1}^{3} F_{i0}(\omega_i)\frac{\partial \omega_i}{\partial x_j}\frac{\partial \omega_i}{\partial x_j} + i\frac{\partial Q}{\partial t} + \Delta Q + 2ieA_a\frac{\partial Q}{\partial x_a}$$
$$+Q\left(T_0(t) + ie\frac{\partial A_a}{\partial x_a} - eA_0 - e^2 A_a A_a\right) = 0. \qquad (15)$$

Thus, the problem of variable separation in SE reduces to integrating the system of ten nonlinear PDEs for four functions. What is more, some coefficients are arbitrary functions that should be determined while integrating equations (12)–(15). We have succeeded in constructing the general solution of the later, which yields, in particular, all possible vector potentials $A(t, \vec{x}) = (A_0(t, \vec{x}), \ldots, A_3(t, \vec{x}))$ such that SE (1) is solvable by the method of the separation of variables.

The integration procedure relies heavily upon the results on the separation of variables in the stationary SE (2). That is why we will briefly consider the principal steps of application of our approach for separating variables in equation (2) with $E = 1$,
$$\Delta_3 \psi - \psi = 0. \qquad (16)$$

Inserting the separation Ansätz,
$$\psi(\vec{x}) = Q(\vec{x})\varphi_1(\omega_1(\vec{x}))\varphi_2(\omega_2(\vec{x}))\varphi_3(\omega_3(\vec{x})),$$

into (16) and taking into account the relations
$$\varphi_a'' = (F_{a1}(\omega_a) + F_{a2}(\omega_a)\lambda_1 + F_{a3}(\omega_a)\lambda_2)\,\varphi_a, \quad a = 1, 2, 3,$$

we get the system of nonlinear partial differential equations for the functions $Q, \omega_1, \omega_2, \omega_3$,

$$\begin{aligned}
(i) & \quad \frac{\partial \omega_i}{\partial x_a}\frac{\partial \omega_j}{\partial x_a} = 0, \quad i \neq j, \quad i,j = 1,2,3; \\
(ii) & \quad \sum_{i=1}^{3} F_{ia}(\omega_i)\frac{\partial \omega_i}{\partial x_j}\frac{\partial \omega_i}{\partial x_j} = 0, \quad a = 2,3; \\
(iii) & \quad 2\frac{\partial Q}{\partial x_j}\frac{\partial \omega_a}{\partial x_j} + Q\Delta\omega_a = 0, \quad a = 1,2,3; \qquad (17)\\
(iv) & \quad Q\sum_{i=1}^{3} F_{i1}(\omega_i)\frac{\partial \omega_i}{\partial x_j}\frac{\partial \omega_i}{\partial x_j} + \Delta Q - Q = 0.
\end{aligned}$$

Now we can utilize the classical results on variable separation in the stationary SE (16). According to [8] the general solution $\vec{\omega} = \vec{\omega}(\vec{x})$ of system (17) splits into eleven inequivalent classes whose representatives are given in (4).

As $\omega_1, \omega_2, \omega_3$ are functionally independent, the inequality

$$\det \left\| \frac{\partial \omega_i}{\partial x_a} \right\|_{i,a=1}^{3} \neq 0 \tag{18}$$

holds. Taking into account this fact and relations (5), we get from $iii$ that $Q(\vec{x}) = \text{const}$, whence it follows that without losing generality we may choose $Q(\vec{x}) = 1$. In view of this we rewrite system (17) in the following way:

$$\frac{\partial \omega_i}{\partial x_a} \frac{\partial \omega_j}{\partial x_a} = 0, \quad i \neq j, \quad i, j = 1, 2, 3;$$
$$\sum_{i=1}^{3} F_{ij}(\omega_i) \omega_{ix_a} \omega_{ix_a} = \delta_{1j}, \quad j = 1, 2, 3, \tag{19}$$

where $\delta_{1j}$ is the Kronecker symbol.

System (19) coincides with the system of equations (12) and (13) under $T_1 = 1, T_2 = T_3 = 0$. Next, in view of arbitrariness of the choice of the separation constants $\lambda_1, \lambda_2, \lambda_3$ we can always suppose that $T_1(t_0) = 1, T_1(t_0) = T_2(t_0) = 0$ for some $t_0 \in \mathbf{R}$. Consequently, system (12), (13) with $t = t_0$ takes the form (19). This is a key point enabling us to use the results of [8] for integrating system (12), (13).

**Lemma 1** *The general solution $\vec{\omega} = \vec{\omega}(t, \vec{x})$ of the system of partial differential equations (12), (13) is given implicitly by the following formulas:*

$$\vec{x} = \mathcal{T}(t) H(t) \vec{z}(\vec{\omega}) + \vec{w}(t). \tag{20}$$

*Here $\mathcal{T}(t)$ is the time-dependent $3 \times 3$ orthogonal matrix:*

$$\mathcal{T}(t) = \begin{pmatrix} \cos\alpha \cos\beta - \sin\alpha \sin\beta \cos\gamma & -\cos\alpha \sin\beta - \sin\alpha \cos\beta \cos\gamma & \sin\alpha \sin\gamma \\ \sin\alpha \cos\beta + \cos\alpha \sin\beta \cos\gamma & -\sin\alpha \sin\beta + \cos\alpha \cos\beta \cos\gamma & -\cos\alpha \sin\gamma \\ \sin\beta \sin\gamma & \cos\beta \sin\gamma & \cos\gamma \end{pmatrix}, \tag{21}$$

$\alpha, \beta, \gamma$ being arbitrary smooth functions of $t$; $\vec{z} = \vec{z}(\vec{\omega})$ is given by one of the eleven formulas from (4); $H(t)$ is the $3 \times 3$ diagonal matrix,

$$H(t) = \begin{pmatrix} h_1(t) & 0 & 0 \\ 0 & h_2(t) & 0 \\ 0 & 0 & h_3(t) \end{pmatrix}, \qquad (22)$$

where

(a) $h_1(t), h_2(t), h_3(t)$ are arbitrary smooth functions for the completely split coordinate system (case 1 from (4)),

(b) $h_1(t) = h_2(t)$, $h_1(t), h_3(t)$ being arbitrary smooth functions, for the partially split coordinate systems (cases 2–4 from (4)),

(c) $h_1(t) = h_2(t) = h_3(t)$, $h_1(t)$ being an arbitrary smooth function, for non-split coordinate systems (cases 5–11 from (4))

and $\vec{w}(t)$ stands for the vector-column whose entries $w_1(t), w_2(t), w_3(t)$ are arbitrary smooth functions of $t$.

**Proof.** First we perform the hodograph transformation in the system of PDEs (12), (13),

$$t = t, \quad x_a = u_a(t, \omega_1, \omega_2, \omega_3), \quad a = 1, 2, 3. \qquad (23)$$

Direct computation shows that the following identities hold:

1. $\omega_{ix_a}\omega_{jx_a} \equiv \dfrac{1}{\delta^2}(\Omega_{ik}\Omega_{jk} - \Omega_{ij}\Omega_{kk}), \quad (i,j,k) = \text{cycle}\,(1,2,3),$

2. $\omega_{ix_a}\omega_{ix_a} \equiv \dfrac{1}{\delta^2}(\Omega_{jj}\Omega_{kk} - \Omega_{jk}^2), \quad (i,j,k) = \text{cycle}\,(1,2,3), \qquad (24)$

where

$$\Omega_{ij} = \frac{\partial u_a}{\partial \omega_i}\frac{\partial u_a}{\partial \omega_j}, \quad \delta = \det\left\|\frac{\partial u_a}{\partial \omega_b}\right\|_{a,b=1}^3 \neq 0.$$

Consequently, the initial system (12), (13), after being rewritten in the new variables, reads as

$$\frac{\partial u_a}{\partial \omega_i}\frac{\partial u_a}{\partial \omega_j} = 0, \quad i \neq j, \quad i,j = 1,2,3;$$

$$\sum_{(i,j,k)=\text{cycle}\,(1,2,3)} F_{ia}(\omega_i)\frac{\partial u_b}{\partial \omega_j}\frac{\partial u_b}{\partial \omega_j}\frac{\partial u_c}{\partial \omega_k}\frac{\partial u_c}{\partial \omega_k} = \delta^2 T_a(t), \quad a = 1,2,3. \qquad (25)$$

If we introduce three vectors,

$$\vec{v}_1 = \frac{\partial \vec{u}}{\partial \omega_1}, \quad \vec{v}_2 = \frac{\partial \vec{u}}{\partial \omega_2}, \quad \vec{v}_3 = \frac{\partial \vec{u}}{\partial \omega_3},$$

then the first three equations of system (25) read as $\vec{v}_a \vec{v}_b = 0$, $(a, b = 1, 2, 3, a \neq b)$. Consequently, $\vec{v}_1, \vec{v}_2, \vec{v}_3$ form an orthogonal system of vectors in the space $\mathbf{R}^3$. It is well known from the analytical geometry that the most general form of these vectors can be expressed via the Euler angles,

$$\begin{aligned}
\frac{\partial \vec{u}}{\partial \omega_1} &= R_1 \left\{ \begin{array}{c} \cos f_1 \cos f_2 - \sin f_1 \sin f_2 \cos f_3 \\ \sin f_1 \cos f_2 + \cos f_1 \sin f_2 \cos f_3 \\ \sin f_2 \sin f_3 \end{array} \right\}, \\
\frac{\partial \vec{u}}{\partial \omega_2} &= R_2 \left\{ \begin{array}{c} -\cos f_1 \sin f_2 - \sin f_1 \cos f_2 \cos f_3 \\ -\sin f_1 \sin f_2 + \cos f_1 \cos f_2 \cos f_3 \\ \cos f_2 \sin f_3 \end{array} \right\}, \quad (26) \\
\frac{\partial \vec{u}}{\partial \omega_3} &= R_3 \left\{ \begin{array}{c} \sin f_1 \sin f_3 \\ -\cos f_1 \sin f_3 \\ \cos f_3 \end{array} \right\},
\end{aligned}$$

where $f_1, \ldots, R_3$ are arbitrary smooth functions of $t, \omega_1, \omega_2, \omega_3$. The above formulas give the most general form of functions $\partial u_a/\partial \omega_b$, $a, b = 1, 2, 3$ that satisfy the first three equations from (25). Next, inserting (26) into the remaining equations from (25) yields

$$\sum_{i=1}^{3} F_{ij}(\omega_i) R_i^{-2} = T_j(t), \quad j = 1, 2, 3. \tag{27}$$

Thus, we have transformed system (25) to equivalent form (26), (27), where one should take into account the compatibility conditions for the system of PDEs (26),

$$\frac{\partial}{\partial \omega_k}\left(\frac{\partial u_i}{\partial \omega_j}\right) = \frac{\partial}{\partial \omega_j}\left(\frac{\partial u_i}{\partial \omega_k}\right), \quad j \neq k, \quad i, j, k = 1, 2, 3.$$

Hence, we get the overdetermined system of nonlinear PDEs for the functions $f_1, f_2, f_3$,

$$\cos f_3 \frac{\partial f_1}{\partial \omega_1} + \frac{\partial f_2}{\partial \omega_1} = -R_2^{-1} \frac{\partial R_1}{\partial \omega_2},$$

$$\cos f_3 \frac{\partial f_1}{\partial \omega_2} + \frac{\partial f_2}{\partial \omega_2} = R_1^{-1} \frac{\partial R_2}{\partial \omega_1},$$

$$\cos f_3 \frac{\partial f_1}{\partial \omega_3} + \frac{\partial f_2}{\partial \omega_3} = 0,$$

$$\cos f_2 \frac{\partial f_3}{\partial \omega_1} + \sin f_2 \sin f_3 \frac{\partial f_1}{\partial \omega_1} = 0,$$

$$\cos f_2 \frac{\partial f_3}{\partial \omega_2} + \sin f_2 \sin f_3 \frac{\partial f_1}{\partial \omega_2} = -R_3^{-1} \frac{\partial R_2}{\partial \omega_3},$$

$$\cos f_2 \frac{\partial f_3}{\partial \omega_3} + \sin f_2 \sin f_3 \frac{\partial f_1}{\partial \omega_3} = R_2^{-1} \frac{\partial R_3}{\partial \omega_2},$$

$$\sin f_2 \frac{\partial f_3}{\partial \omega_1} - \cos f_2 \sin f_3 \frac{\partial f_1}{\partial \omega_1} = -R_3^{-1} \frac{\partial R_1}{\partial \omega_3},$$

$$\sin f_2 \frac{\partial f_3}{\partial \omega_2} - \cos f_2 \sin f_3 \frac{\partial f_1}{\partial \omega_2} = 0,$$

$$\sin f_2 \frac{\partial f_3}{\partial \omega_3} - \cos f_2 \sin f_3 \frac{\partial f_1}{\partial \omega_3} = R_1^{-1} \frac{\partial R_3}{\partial \omega_1}.$$

While integrating the above system we have to differentiate between the two cases: (1) $\sin f_3 \neq 0$ and (2) $\sin f_3 = 0$.

**Case 1.** Suppose the condition $\sin f_3 \neq 0$ holds. Then we can solve the above system with respect to $\partial f_a / \partial \omega_b$, $(a, b = 1, 2, 3)$ and get

$$\begin{aligned}
\frac{\partial f_1}{\partial \omega_1} &= R_3^{-1} \frac{\partial R_1}{\partial \omega_3} \cos f_2 \csc f_3, \\
\frac{\partial f_1}{\partial \omega_2} &= -R_3^{-1} \frac{\partial R_2}{\partial \omega_3} \sin f_2 \csc f_3, \\
\frac{\partial f_1}{\partial \omega_3} &= R_2^{-1} \frac{\partial R_3}{\partial \omega_2} \sin f_2 \csc f_3 - R_1^{-1} \frac{\partial R_3}{\partial \omega_1} \cos f_2 \csc f_3, \\
\frac{\partial f_2}{\partial \omega_1} &= -R_2^{-1} \frac{\partial R_1}{\partial \omega_2} - R_3^{-1} \frac{\partial R_1}{\partial \omega_3} \cos f_2 \cot f_3, \\
\frac{\partial f_2}{\partial \omega_2} &= R_1^{-1} \frac{\partial R_2}{\partial \omega_1} + R_3^{-1} \frac{\partial R_2}{\partial \omega_3} \sin f_2 \cot f_3, \quad (28) \\
\frac{\partial f_2}{\partial \omega_3} &= -R_2^{-1} \frac{\partial R_3}{\partial \omega_2} \sin f_2 \cot f_3 + R_1^{-1} \frac{\partial R_3}{\partial \omega_1} \cos f_2 \cot f_3, \\
\frac{\partial f_3}{\partial \omega_1} &= -R_3^{-1} \frac{\partial R_1}{\partial \omega_3} \sin f_2,
\end{aligned}$$

$$\frac{\partial f_3}{\partial \omega_2} = -R_3^{-1} \frac{\partial R_2}{\partial \omega_3} \cos f_2,$$

$$\frac{\partial f_3}{\partial \omega_3} = R_2^{-1} \frac{\partial R_3}{\partial \omega_2} \cos f_2 + R_1^{-1} \frac{\partial R_3}{\partial \omega_1} \sin f_2.$$

From the compatibility conditions of the above system of PDEs we get the system of nonlinear differential equations for $R_1, R_2, R_3$:

$$
\begin{aligned}
&1.\quad R_1 R_2 \frac{\partial^2 R_3}{\partial \omega_1 \partial \omega_2} - R_1 \frac{\partial R_2}{\partial \omega_1} \frac{\partial R_3}{\partial \omega_2} - R_2 \frac{\partial R_1}{\partial \omega_2} \frac{\partial R_3}{\partial \omega_1} = 0, \\
&2.\quad R_2 R_3 \frac{\partial^2 R_1}{\partial \omega_2 \partial \omega_3} - R_2 \frac{\partial R_1}{\partial \omega_3} \frac{\partial R_3}{\partial \omega_2} - R_3 \frac{\partial R_1}{\partial \omega_2} \frac{\partial R_2}{\partial \omega_3} = 0, \\
&3.\quad R_1 R_3 \frac{\partial^2 R_2}{\partial \omega_1 \partial \omega_3} - R_3 \frac{\partial R_1}{\partial \omega_3} \frac{\partial R_2}{\partial \omega_1} - R_1 \frac{\partial R_2}{\partial \omega_3} \frac{\partial R_3}{\partial \omega_1} = 0, \\
&4.\quad R_1^2 R_2^2 \frac{\partial R_2}{\partial \omega_3} \frac{\partial R_3}{\partial \omega_3} + R_1^2 R_3^2 \frac{\partial R_2}{\partial \omega_2} \frac{\partial R_3}{\partial \omega_2} - R_2^2 R_3^2 \frac{\partial R_2}{\partial \omega_1} \frac{\partial R_3}{\partial \omega_1} \\
&\qquad - R_1^2 R_2^2 R_3 \frac{\partial^2 R_2}{\partial \omega_3 \partial \omega_3} - R_1^2 R_2 R_3^2 \frac{\partial^2 R_3}{\partial \omega_2 \partial \omega_2} = 0, \\
&5.\quad R_1^2 R_2^2 \frac{\partial R_1}{\partial \omega_3} \frac{\partial R_3}{\partial \omega_3} - R_1^2 R_3^2 \frac{\partial R_1}{\partial \omega_2} \frac{\partial R_3}{\partial \omega_2} + R_2^2 R_3^2 \frac{\partial R_1}{\partial \omega_1} \frac{\partial R_3}{\partial \omega_1} \\
&\qquad - R_1^2 R_2^2 R_3 \frac{\partial^2 R_1}{\partial \omega_3 \partial \omega_3} - R_1 R_2^2 R_3^2 \frac{\partial^2 R_3}{\partial \omega_1 \partial \omega_1} = 0, \\
&6.\quad -R_1^2 R_2^2 \frac{\partial R_1}{\partial \omega_3} \frac{\partial R_2}{\partial \omega_3} + R_1^2 R_3^2 \frac{\partial R_1}{\partial \omega_2} \frac{\partial R_2}{\partial \omega_2} + R_2^2 R_3^2 \frac{\partial R_1}{\partial \omega_1} \frac{\partial R_2}{\partial \omega_1} \\
&\qquad - R_1^2 R_2 R_3^2 \frac{\partial^2 R_1}{\partial \omega_2 \partial \omega_2} - R_1 R_2^2 R_3^2 \frac{\partial^2 R_2}{\partial \omega_1 \partial \omega_1} = 0.
\end{aligned}
\quad (29)
$$

Remarkably, there no need for a direct integrating of the system of nonlinear PDEs (29), since it has been solved by Eisenhart [8] under the assumption that $R_1, R_2, R_3$ are independent of $t$. The only thing to be done is to find out in which way the temporal variable $t$ enters into the general solution of system (26)–(29) in the case under study.

As shown above, the problem of variable separation in SE (16) reduces to integrating system (19) and, what is more, the general solution of the latter is given within the equivalence relation (9) by one of the formulas from (4). Consequently, if we fix the temporal variable $t$ to be equal to $t_0 \in \mathbf{R}$, then after integrating equations (26)–(29) we get within the equivalence relation (9) formulas (4).

In view of this fact we can solve relations (25) under $t = t_0$ with respect to $F_{ij}(\omega_i)$ (note that $F_{ij}$ are independent of $t$) for each class of functions $\vec{x} = \vec{z}(\vec{\omega})$ given in (4). The results of these calculations are presented below in the form of $3 \times 3$ Stäckel matrices $\mathcal{F}_1, \ldots, \mathcal{F}_{11}$, whose $(i,j)$th entry is the corresponding function $F_{ij}(\omega_i)$. We give the canonical forms of the matrices $\mathcal{F}_1, \ldots, \mathcal{F}_{11}$ up to the choice of separation constants $\lambda_i$, $i = 1, 2, 3$ in (11).

$$\mathcal{F}_1 = \begin{pmatrix} 1 & 0 & 0 \\ 0 & 1 & 0 \\ 0 & 0 & 1 \end{pmatrix}, \quad \mathcal{F}_2 = \begin{pmatrix} e^{2\omega_1} & -1 & 0 \\ 0 & 1 & 0 \\ 0 & 0 & 1 \end{pmatrix},$$

$$\mathcal{F}_3 = \begin{pmatrix} \omega_1^2 & -1 & 0 \\ \omega_2^2 & 1 & 0 \\ 0 & 0 & 1 \end{pmatrix}, \quad \mathcal{F}_4 = \begin{pmatrix} a^2 \cosh^2 \omega_1 & 1 & 0 \\ -a^2 \cos^2 \omega_2 & -1 & 0 \\ 0 & 0 & 1 \end{pmatrix},$$

$$\mathcal{F}_5 = \begin{pmatrix} \omega_1^{-4} & -\omega_1^{-2} & 0 \\ 0 & \cosh^{-2} \omega_2 & -1 \\ 0 & 0 & 1 \end{pmatrix},$$

$$\mathcal{F}_6 = \begin{pmatrix} a^2 \sinh^{-4} \omega_1 & -\sinh^{-2} \omega_1 & -1 \\ a^2 \cosh^{-4} \omega_2 & \cosh^{-2} \omega_2 & -1 \\ 0 & 0 & 1 \end{pmatrix},$$

$$\mathcal{F}_7 = \begin{pmatrix} a^2 \sin^{-4} \omega_1 & -\sin^{-2} \omega_1 & 1 \\ -a^2 \cosh^{-4} \omega_2 & \cosh^{-2} \omega_2 & -1 \\ 0 & 0 & 1 \end{pmatrix}, \quad (30)$$

$$\mathcal{F}_8 = \begin{pmatrix} e^{4\omega_1} & -e^{2\omega_1} & -1 \\ e^{4\omega_2} & e^{2\omega_2} & -1 \\ 0 & 0 & 1 \end{pmatrix},$$

$$\mathcal{F}_9 = \begin{pmatrix} a^2 \cosh^2 2\omega_1 & -a \cosh 2\omega_1 & -1 \\ -a^2 \cos^2 2\omega_2 & a \cos 2\omega_2 & 1 \\ a^2 \cosh^2 2\omega_3 & a \cosh 2\omega_3 & -1 \end{pmatrix},$$

$$\mathcal{F}_{10} = \begin{pmatrix} \dfrac{\mathrm{dn}^4(\omega_1,k)}{\mathrm{sn}^4(\omega_1,k)} & -\dfrac{\mathrm{dn}^2(\omega_1,k)}{\mathrm{sn}^2(\omega_1,k)} & 1 \\ -k'^4\,\mathrm{cn}^4(\omega_2,k') & k'^2\,\mathrm{cn}^2(\omega_2,k') & -1 \\ k^4\,\mathrm{cn}^4(\omega_3,k) & k^2\,\mathrm{cn}^2(\omega_3,k) & 1 \end{pmatrix},$$

$$\mathcal{F}_{11} = \begin{pmatrix} \omega_1^{-4} & -\omega_1^{-2} & 0 \\ 0 & k'^2\mathrm{cn}^2(\omega_2,k') & -1 \\ 0 & k^2\mathrm{cn}^2(\omega_3,k) & 1 \end{pmatrix}.$$

With the explicit forms of the functions $F_{ij}$, $(i,j=1,2,3)$ in hand, we can solve (27) with respect to $R_1, R_2, R_3$. Inserting the result obtained into (29) and splitting by $\omega_1, \omega_2, \omega_3$ yield the final forms of the functions $R_1, R_2, R_3$:

1. $R_i^2 = T_i^{-1}$, $i = 1,2,3$;
2. $R_1^2 = R_2^2 = T_1^{-1} e^{2\omega_1}$, $R_3^2 = T_3^{-1}$;
3. $R_1^2 = R_2^2 = T_1^{-1}(\omega_1^2 + \omega_2^2)$, $R_3^2 = T_3^{-1}$;
4. $R_1^2 = R_2^2 = T_1^{-1} a^2 (\cosh 2\omega_1 - \cos 2\omega_2)$, $R_3^2 = T_3^{-1}$;
5. $R_1^2 = T_1^{-1} \omega_1^{-4}$, $R_2^2 = R_3^2 = T_1^{-1} \omega_1^{-2} \cosh^{-2} \omega_2$;
6. $R_1^2 = T_1^{-1} a^2 \sinh^{-2}\omega_1 (\sinh^{-2}\omega_1 + \cosh^{-2}\omega_2)$,
   $R_2^2 = T_1^{-1} a^2 \cosh^{-2}\omega_2 (\sinh^{-2}\omega_1 + \cosh^{-2}\omega_2)$,
   $R_3^2 = T_1^{-1} a^2 \sinh^{-2}\omega_1 \cosh^{-2}\omega_2$;
7. $R_1^2 = T_1^{-1} a^2 \sin^{-2}\omega_1 (\sin^{-2}\omega_1 - \cosh^{-2}\omega_2)$,
   $R_2^2 = T_1^{-1} a^2 \cosh^{-2}\omega_2 (\sin^{-2}\omega_1 - \cosh^{-2}\omega_2)$, \hfill (31)
   $R_3^2 = T_1^{-1} a^2 \sin^{-2}\omega_1 \cosh^{-2}\omega_2$;
8. $R_1^2 = T_1^{-1} e^{2\omega_1}(e^{2\omega_1} + e^{2\omega_2})$, $R_2^2 = T_1^{-1} e^{2\omega_2}(e^{2\omega_1} + e^{2\omega_2})$,
   $R_3^2 = T_1^{-1} e^{2(\omega_1 + \omega_2)}$;
9. $R_1^2 = T_1^{-1} a^2 (\cosh 2\omega_1 - \cos 2\omega_2)(\cosh 2\omega_1 + \cosh 2\omega_3)$,
   $R_2^2 = T_1^{-1} a^2 (\cosh 2\omega_1 - \cos 2\omega_2)(\cos 2\omega_2 + \cosh 2\omega_3)$,
   $R_3^2 = T_1^{-1} a^2 (\cosh 2\omega_1 + \cosh 2\omega_3)(\cos 2\omega_2 + \cosh 2\omega_3)$;
10. $R_1^2 = T_1^{-1} \left( \dfrac{\mathrm{dn}^2(\omega_1,k)}{\mathrm{sn}^2(\omega_1,k)} - k'^2 \mathrm{cn}^2(\omega_2,k') \right) \left( \dfrac{\mathrm{dn}^2(\omega_1,k)}{\mathrm{sn}^2(\omega_1,k)} + k^2 \mathrm{cn}^2(\omega_3,k) \right)$,
    $R_2^2 = T_1^{-1} \left( \dfrac{\mathrm{dn}^2(\omega_1,k)}{\mathrm{sn}^2(\omega_1,k)} - k'^2 \mathrm{cn}^2(\omega_2,k') \right) \left( k'^2 \mathrm{cn}^2(\omega_2,k') + k^2 \mathrm{cn}^2(\omega_3,k) \right)$,

$$R_3^2 = T_1^{-1} \left( \frac{\mathrm{dn}^2(\omega_1, k)}{\mathrm{sn}^2(\omega_1, k)} + k^2 \mathrm{cn}^2(\omega_3, k) \right) \left( k'^2 \mathrm{cn}^2(\omega_2, k') + k^2 \mathrm{cn}^2(\omega_3, k) \right);$$

11. $R_1^2 = T_1^{-1} \omega_1^{-4}, \quad R_2^2 = R_3^2 = T_1^{-1} \omega_1^{-2} \left( k'^2 \mathrm{cn}^2(\omega_2, k') + k^2 \mathrm{cn}^2(\omega_3, k) \right).$

Inserting the above expressions into (28), we see that the system obtained (we denote it temporarily as $\mathcal{S}$) does not depend explicitly on $t$. It is in involution and hence its general solution depends on three arbitrary functions of $t$. As a direct check shows the functions

$$\begin{aligned}
f_1 &= \operatorname{arccot}\left( \cos\gamma \cot(\tilde{f}_1 + \beta) + \sin\gamma \cot\tilde{f}_3 \csc(\tilde{f}_1 + \beta) \right) + \alpha, \\
f_2 &= \operatorname{arccot}\left( \cos\tilde{f}_3 \cot(\tilde{f}_1 + \beta) + \sin\tilde{f}_3 \cot\gamma \csc(\tilde{f}_1 + \beta) \right) + \tilde{f}_2, \quad (32) \\
f_3 &= \arccos\left( \cos\gamma \cos\tilde{f}_3 - \sin\gamma \sin\tilde{f}_3 \cos(\tilde{f}_1 + \beta) \right),
\end{aligned}$$

where $\alpha, \beta, \gamma$ are arbitrary functions of $t$ and $\tilde{f}_1, \tilde{f}_2, \tilde{f}_3$ are (time-independent) solutions of the system $\mathcal{S}$ under $t = t_0$, satisfy the system $\mathcal{S}$ identically. Since formulas (32) contain three arbitrary functions of $t$, they give the general solution of the system $\mathcal{S}$. Substituting (32) into (26) yields

$$\begin{aligned}
\frac{\partial \vec{u}}{\partial \omega_1} &= R_1 \mathcal{T}(t) \left\{ \begin{array}{c} \cos\tilde{f}_1 \cos\tilde{f}_2 - \sin\tilde{f}_1 \sin\tilde{f}_2 \cos\tilde{f}_3 \\ \sin\tilde{f}_1 \cos\tilde{f}_2 + \cos\tilde{f}_1 \sin\tilde{f}_2 \cos\tilde{f}_3 \\ \sin\tilde{f}_2 \sin\tilde{f}_3 \end{array} \right\}, \\
\frac{\partial \vec{u}}{\partial \omega_2} &= R_2 \mathcal{T}(t) \left\{ \begin{array}{c} -\cos\tilde{f}_1 \sin\tilde{f}_2 - \sin\tilde{f}_1 \cos\tilde{f}_2 \cos\tilde{f}_3 \\ -\sin\tilde{f}_1 \sin\tilde{f}_2 + \cos\tilde{f}_1 \cos\tilde{f}_2 \cos\tilde{f}_3 \\ \cos\tilde{f}_2 \sin\tilde{f}_3 \end{array} \right\}, \quad (33) \\
\frac{\partial \vec{u}}{\partial \omega_3} &= R_3 \mathcal{T}(t) \left\{ \begin{array}{c} \sin\tilde{f}_1 \sin\tilde{f}_3 \\ -\cos\tilde{f}_1 \sin\tilde{f}_3 \\ \cos\tilde{f}_3 \end{array} \right\},
\end{aligned}$$

where the matrix $\mathcal{T}(t)$ is given by formula (21).

If we choose in the system obtained $t = t_0$, then its general solution is given within the equivalence relation $\mathcal{E}$ (9) by one of the formulas from (4). In view of this fact it is not difficult to integrate system (33) and thus get formulas (20), where $h_1(t), h_2(t), h_3(t)$ are expressed via $T_1(t), T_2(t), T_3(t)$,

$$\begin{aligned}
1. \quad & T_i = h_i^{-2}, \quad i = 1, 2, 3; \\
2-4. \quad & T_1 = h_1^{-2}, \quad T_2 = 0, \quad T_3 = h_3^{-2}; \quad (34) \\
5-11. \quad & T_1 = h_1^{-2}, \quad T_2 = T_3 = 0.
\end{aligned}$$

**Case 2.** Let the relation $\sin f_3 = 0$ be valid. In this case we have an analog of system (28),

$$\frac{\partial g}{\partial \omega_1} = -R_2^{-1}\frac{\partial R_1}{\partial \omega_2}, \quad \frac{\partial g}{\partial \omega_2} = R_1^{-1}\frac{\partial R_2}{\partial \omega_1},$$
$$\frac{\partial g}{\partial \omega_3} = 0, \quad \frac{\partial R_1}{\partial \omega_3} = 0, \quad \frac{\partial R_2}{\partial \omega_3} = 0, \quad \frac{\partial R_3}{\partial \omega_1} = 0, \quad \frac{\partial R_3}{\partial \omega_2} = 0, \quad (35)$$

with $g = \pm f_1 + f_2$ and an analog of system (29),

$$R_1^2 R_3^2 \frac{\partial R_1}{\partial \omega_2}\frac{\partial R_2}{\partial \omega_2} + R_2^2 R_3^2 \frac{\partial R_1}{\partial \omega_1}\frac{\partial R_2}{\partial \omega_1} - R_1^2 R_2 R_3^2 \frac{\partial^2 R_1}{\partial \omega_2 \partial \omega_2}$$
$$- R_1 R_2^2 R_3^2 \frac{\partial^2 R_2}{\partial \omega_1 \partial \omega_1} = 0, \quad (36)$$
$$\frac{\partial R_1}{\partial \omega_3} = 0, \quad \frac{\partial R_2}{\partial \omega_3} = 0, \quad \frac{\partial R_3}{\partial \omega_1} = 0, \quad \frac{\partial R_3}{\partial \omega_2} = 0.$$

System of PDEs (35), (36) is fairly simple and is easily integrated. As a result we get a particular case of (20) with $\sin \gamma = 0$. The lemma is proved. ▷

With this result in hand it is not difficult to integrate the remaining equations from the system under study. Indeed, equations (14) and (15) may be treated as algebraic equations for the functions $A_j(t,\vec{x})$, $j = 1,2,3$ and $A_0(t,\vec{x})$, correspondingly.

There are two different configurations of the electromagnetic field that should be considered separately. The first one is the case of a vanishing magnetic field $\vec{H} = \mathrm{rot}\,\vec{A}$, namely, the case when

$$A_{i x_j} = A_{j x_i}, \quad i \neq j, \quad i,j = 1,2,3. \quad (37)$$

Provided the above equality does not hold we have a nonvanishing magnetic field $\vec{H}$.

**1. The case of nonvanishing magnetic field.**

Let us represent the complex-valued function $Q$ in (6) as $Q = \exp(S_1 + iS_2)$, where $S_1, S_2$ are real-valued functions. Now, if we take into account that the components of the vector potential $A(t,\vec{x})$ and functions $\omega_1, \omega_2, \omega_3$ are real-valued functions, then after inserting $Q$ into (14) with the use of (5) we can split the obtained equations into real and imaginary parts:

$$\frac{\partial S_1}{\partial x_j}\frac{\partial \omega_a}{\partial x_j} = 0, \quad a = 1, 2, 3; \tag{38}$$

$$2\left(\frac{\partial S_2}{\partial x_j} + eA_j\right)\frac{\partial \omega_a}{\partial x_j} + \frac{\partial \omega_a}{\partial t} = 0, \quad a = 1, 2, 3. \tag{39}$$

In view of (18) we get from (38) the relations $S_{1x_j} = 0$, $j = 1, 2, 3$. Hence, $S_1 = S_1(t)$ and within the equivalence relation $\mathcal{E}$ (10) we may choose $S_1$ to be equal to 0. Next, making use of the gauge invariance of SE (1) we get from (39) that within the equivalence relation $\mathcal{E}$ the equalities $S_{2x_i} = 0$, $i = 1, 2, 3$ hold, whence $S_2 = S_2(t)$. Again, in view of (10) we may put $S_2 = 0$, which means that the factor $Q$ may be chosen as 1. With this result system (39) reduces to the system of three linear algebraic equations for the functions $A_1$, $A_2$, $A_3$,

$$\omega_{at} = -2e\omega_{ax_i}A_i, \quad a = 1, 2, 3.$$

The determinant of this system is not equal to zero due to (18). Consequently, it has a unique solution. Making in this solution the hodograph transformation (23), we get the following expressions for $A_1, A_2, A_3$:

$$\vec{A} = \frac{1}{2e}\frac{\partial \vec{u}(t,\vec{\omega})}{\partial t}.$$

Inserting into the above formula $\vec{x} = \vec{u}(t,\vec{\omega})$ from (20) yields the explicit forms of the spacelike components of the vector potential of the electromagnetic field $A(t,\vec{x})$,

$$\vec{A}(t,\vec{x}) = \frac{1}{2e}\left(\mathcal{M}(t)(\vec{x}-\vec{w}) + \dot{\vec{w}}\right). \tag{40}$$

Here we use the designation

$$\mathcal{M}(t) = \dot{\mathcal{T}}(t)\mathcal{T}^{-1}(t) + \mathcal{T}(t)\dot{H}(t)H^{-1}(t)\mathcal{T}^{-1}(t), \tag{41}$$

where $\mathcal{T}(t)$, $H(t)$ are variable $3 \times 3$ matrices defined by formulas (21) and (22), correspondingly, $\vec{w} = (w_1(t), w_2(t), w_3(t))^T$ and the dot over a symbol means differentiation with respect to $t$.

Given the form (40) of the spacelike components of the vector potential $A(t, \vec{x})$, the condition for the magnetic field $\vec{H}$ not to vanish means that at least one of the expressions

$$\dot{\alpha} + \dot{\beta}\cos\gamma, \quad \dot{\beta}\cos\alpha\sin\gamma - \dot{\gamma}\sin\alpha, \quad \dot{\beta}\sin\alpha\sin\gamma + \dot{\gamma}\cos\alpha \qquad (42)$$

does not turn into zero. Hence, in view of the identity

$$\dot{\mathcal{T}}\mathcal{T}^{-1} = \begin{pmatrix} 0 & -(\dot{\alpha} + \dot{\beta}\cos\gamma) \\ \dot{\alpha} + \dot{\beta}\cos\gamma & 0 \\ \dot{\beta}\cos\alpha\sin\gamma - \dot{\gamma}\sin\alpha & \dot{\beta}\sin\alpha\sin\gamma + \dot{\gamma}\cos\alpha \end{pmatrix} \rightarrow$$

$$\rightarrow \begin{pmatrix} -(\dot{\beta}\cos\alpha\sin\gamma - \dot{\gamma}\sin\alpha) \\ -(\dot{\beta}\sin\alpha\sin\gamma + \dot{\gamma}\cos\alpha) \\ 0 \end{pmatrix},$$

whose validity is checked by straightforward computation, we conclude that $\mathcal{T} \neq$ const. At last, solving (15) yields the explicit form of $A_0$,

$$A_0(t, \vec{x}) = \frac{1}{e}\left(\sum_{i=1}^{3} F_{i0}(\omega_i)\frac{\partial\omega_i}{\partial x_j}\frac{\partial\omega_i}{\partial x_j} + T_0(t)\right) + iA_{ax_a} - eA_aA_a.$$

As the function $A_0$ is real valued, we have to find the real part of the right-hand side of the above equality. Making use of (40) yields

$$2eA_{ax_a} = \sum_{i=1}^{3} \frac{\dot{h}_i}{h_i}.$$

This, in its turn, gives the imaginary part of $T_0 = T_0(t)$,

$$T_0 = \tilde{T}_0 - \frac{i}{2}\sum_{i=1}^{3}\frac{\dot{h}_i}{h_i}, \quad \operatorname{Im}\tilde{T}_0 = 0. \qquad (43)$$

Finally, we get

$$A_0(t, \vec{x}) = \frac{1}{e}\left(\sum_{i=1}^{3} F_{i0}(\omega_i)\frac{\partial\omega_i}{\partial x_j}\frac{\partial\omega_i}{\partial x_j} + \tilde{T}_0(t)\right) - eA_aA_a, \qquad (44)$$

where $F_{10}(\omega_1), F_{20}(\omega_2), F_{30}(\omega_3)$ are arbitrary smooth functions, $A_1, A_2, A_3$ are given by (40), and the functions $\omega_a = \omega_a(t, \vec{x}), a = 1, 2, 3$ are defined implicitly by formulas (20)–(22) and (4). In order to keep an exposition of the results compact, we do not present here $A_0(t, \vec{x})$ in full detail, since the corresponding expressions are too cumbersome.

Thus we have proved the following assertion.

**Theorem 1** *SE (1) for the case of nonvanishing magnetic field admits separation of variables if and only if the spacelike components $A_1, A_2, A_3$ of the vector potential of the electromagnetic field are linear in the spatial variables and given by (40) and, furthermore, the timelike component $A_0$ is given by (44).*

Summing up, we conclude that conditions (40) and (44) provide separability of SE for the case of a nonvanishing magnetic field. And what is more, the solutions with separated variables are of the form (6) with $Q = 1$, where the functions $\omega_1(t, \vec{x}), \omega_2(t, \vec{x}), \omega_3(t, \vec{x})$ are given implicitly by formulas (20)–(22) and (4). In fact, we have the eleven classes of vector potentials $A(t, \vec{x})$ corresponding to the eleven classes of coordinate systems $\omega_a = \omega_a(t, \vec{x})$, $a = 1, 2, 3$. SE (1) for each class of the functions $A_0(t, \vec{x}), \vec{A}(t, \vec{x})$ defined by (40) and (44) under arbitrary $\tilde{T}_0(t), F_{a0}(\omega_a)$ and fixed arbitrary functions $\alpha(t), \beta(t), \gamma(t)$, $w_a(t), h_a(t), a = 1, 2, 3$ separates in exactly one coordinate system. The separation equations read as (11), where the coefficients $F_{ai}, a, i = 1, 2, 3$ are the entries of the corresponding Stäckel matrices (30), functions $T_a, a = 1, 2, 3$ are listed in (34), and the functions $T_0, F_{a0}, a = 1, 2, 3$ are arbitrary smooth functions defining the form of the timelike component of the vector potential $A(t, \vec{x})$ (see (44)).

Note that proper specifying the functions $F_{a0}(\omega_a), a = 1, 2, 3$, may yield additional possibilities for variable separation in the corresponding SE. What we mean is that for some particular forms of the vector potential $A(t, \vec{x})$ (40), (44) there might exist several coordinate systems (20)–(22) enabling us to separate the corresponding SE. However, the detailed study of this problem goes beyond the scope of the present paper.

As an illustration of the previous subsection, we consider the problem of separation of variables in SE (1), where the vector potential of the electro-magnetic field is of the form

$$e\vec{A} = \begin{pmatrix} 0 & -s_1(t) & -s_2(t) \\ s_1(t) & 0 & -s_3(t) \\ s_2(t) & s_3(t) & 0 \end{pmatrix} \vec{x}, \qquad (45)$$

$$eA_0 = \frac{q}{|\vec{x}|} - \left((s_1(t)x_2 + s_2(t)x_3)^2 \right.$$
$$\left. + (s_1(t)x_1 - s_3(t)x_3)^2 + (s_2(t)x_1 + s_3(t)x_2)^2\right).$$

Here $q = \text{const}$ and

$$2s_1(t) = \dot{\alpha}(t) + \dot{\beta}(t)\cos\gamma(t),$$
$$2s_2(t) = \dot{\beta}(t)\cos\alpha(t)\sin\gamma(t) - \dot{\gamma}(t)\sin\alpha(t),$$
$$2s_3(t) = \dot{\beta}(t)\sin\alpha(t)\sin\gamma(t) + \dot{\gamma}(t)\cos\alpha(t),$$

where $\alpha(t), \beta(t), \gamma(t)$ are arbitrary smooth functions. Evidently, choosing $\alpha(t)$=const, $\beta(t)$=const, and $\gamma(t)$=const yields the standard Coulomb potential.

Making use of the results of Theorem 1, we conclude that SE with potential of the form (45) separates in four coordinate systems,

$$\vec{x} = \mathcal{T}(t)\vec{z},$$

where $\mathcal{T}$ is the time-dependent $3 \times 3$ orthogonal matrix (21), and $\vec{z}$ is one of the following coordinate systems:

1. spherical (formula 5 from (4)),

2. prolate spheroidal II (formula 6 from (4), where one should replace $z_3$ with $z_3 = a(\coth\omega_1 \tanh\omega_2 \pm 1)$),

3. parabolic (formula 8 from (4)),

4. conical (formula 11 from (4)).

The separation equations (11) for these cases take the form

1.
$$i\varphi_0' = -\lambda_1\varphi_0,$$
$$\varphi_1'' = (\lambda_1\omega_1^{-4} - \lambda_2\omega_1^{-2} + q\omega_1^{-3})\varphi_1,$$
$$\varphi_2'' = (\lambda_2 \operatorname{sech}^2\omega_2 - \lambda_3)\varphi_2,$$
$$\varphi_3'' = \lambda_3\varphi_3.$$

Integrating these equations yields a family of exact solutions of SE (1) with potential (45) that are products of the exponential, confluent hypergeometric [30], and Legendre [30] functions.

2.
$$i\varphi_0' = -\lambda_1\varphi_0,$$
$$\varphi_1'' = (\lambda_1 a^2 \sinh^{-4}\omega_1 - \lambda_2 \sinh^{-2}\omega_1 - \lambda_3 + qa\cosh\omega_1 \sinh^{-3}\omega_1)\varphi_1,$$
$$\varphi_2'' = (\lambda_1 a^2 \cosh^{-4}\omega_2 + \lambda_2 \cosh^{-2}\omega_2 - \lambda_3 \mp qa\sinh\omega_2 \cosh^{-3}\omega_2)\varphi_2,$$
$$\varphi_3'' = \lambda_3\varphi_3.$$

Integrating these equations yields a family of exact solutions of SE (1) with potential (45) in separated form that are products of the exponential and Coulomb spheroidal functions [31].

3.
$$i\varphi_0' = -\lambda_1\varphi_0,$$
$$\varphi_1'' = (\lambda_1 e^{4\omega_1} - \lambda_2 e^{2\omega_1} - \lambda_3 + 2qe^{2\omega_1})\varphi_1,$$
$$\varphi_2'' = (\lambda_1 e^{4\omega_2} + \lambda_2 e^{2\omega_2} - \lambda_3)\varphi_1,$$
$$\varphi_3'' = \lambda_3\varphi_3.$$

Integrating these equations yields a family of exact solutions of SE (1) with potential (45) that are products of the exponential and confluent hypergeometric functions.

4.
$$i\varphi_0' = -\lambda_1\varphi_0,$$
$$\varphi_1'' = (\lambda_1\omega_1^{-4} - \lambda_2\omega_1^{-2} + q\omega_1^{-3})\varphi_1,$$
$$\varphi_2'' = (\lambda_2 k'^2 \operatorname{cn}^2(\omega_2, k') - \lambda_3)\varphi_2,$$
$$\varphi_3'' = (\lambda_2 k^2 \operatorname{cn}^2(\omega_3, k) + \lambda_3)\varphi_3.$$

Integrating these equations yields a family of exact solutions of SE (1) with potential (45) that are products of the exponential, confluent hypergeometric, and Lamé [30] functions.

Note that these families of solutions are parametrized both by integration constants and by the three continuous spectral parameters $\lambda_i, i = 1, 2, 3$. Under appropriate initial and boundary conditions the latter become discrete, i.e. $\lambda_i = \lambda_{in}, n = 1, 2, 3, \ldots$, and we get a basis for expanding arbitrary smooth solutions of SE (1) with potential (45) in a properly chosen Hilbert space (for more details, see [15]).

## 2. The case of vanishing magnetic field.

Provided condition (37) holds true, we can, without loss of generality, put $\vec{A} = 0$ at the expense of the gauge invariance of SE. Representing the complex-valued function $Q$ in (6) as $Q = \exp(S_1 + iS)$, inserting into (14) and splitting the obtained equations into real and imaginary parts analogously to what has been done for the case of a nonvanishing magnetic field we arrive at the conclusion that $S_1$ may be chosen to be equal to zero. Furthermore, the function $S = S(t, \vec{x})$ satisfy the over-determined system of linear partial differential equations

$$2\frac{\partial S}{\partial x_j}\frac{\partial \omega_a}{\partial x_j} + \frac{\partial \omega_a}{\partial t} = 0, \quad a = 1, 2, 3. \tag{46}$$

Solving it with respect to the derivatives $\partial S/\partial x_j$, $j = 1, 2, 3$ (which is always possible due to (18)) and making in the relations obtained the hodograph transformation (23), yields

$$2\vec{\nabla}S = \mathcal{M}(t)(\vec{x} - \vec{w}) + \dot{\vec{w}}, \tag{47}$$

where $\mathcal{M}(t)$ is the 3×3 matrix (41) and $\vec{w} = (w_1(t), w_2(t), w_3(t))^T$. The direct check shows that system (47) is compatible if and only if all the expressions given in (42) vanish identically. Hence, we conclude that the matrix $\mathcal{T}$ is constant. Utilizing the invariance of SE (1) with respect to the rotation group we can choose $\mathcal{T} = I$, where $I$ is the unit $3 \times 3$ matrix. Given this condition, (20) takes the form

$$\vec{x} = H(t)\vec{z}(\vec{\omega}) + \vec{w}(t), \tag{48}$$

and, furthermore,
$$\mathcal{M}(t) = \dot{H}(t)H^{-1}(t).$$

Next, integrating (47) and taking into account the equivalence relation (10) we get
$$S = \frac{1}{2}\sum_{i=1}^{3}\left(\frac{\dot{h}_i}{h_i}\left(\frac{x_i^2}{2} - w_i x_i\right) + \dot{w}_i x_i\right). \qquad (49)$$

Substituting (49) into (15) yields the form of $A_0(t,\vec{x})$,

$$\begin{aligned}
eA_0(t,\vec{x}) &= \sum_{i=1}^{3} F_{i0}(\omega_i)\frac{\partial \omega_i}{\partial x_j}\frac{\partial \omega_i}{\partial x_j} + \tilde{T}_0(t) \\
&\quad -\frac{1}{4}\sum_{i=1}^{3}\left(\frac{\ddot{h}_i}{h_i}x_i^2 + 2\left(\ddot{w}_i - \frac{\ddot{h}_i}{h_i}w_i\right)x_i + \left(\dot{w}_i - \frac{\dot{h}_i}{h_i}w_i\right)^2\right), \quad (50)
\end{aligned}$$

the function $T_0(t)$ being given by (43).

Thus, we have proved the following assertion.

**Theorem 2** *Given the restriction (37), SE (1) admits a separation of variables if and only if it is gauge equivalent to SE with $\vec{A} = \vec{0}$ and $A_0$ being given by (50).*

Consequently, the conditions $\vec{A} = \vec{0}$ and (50) provide separability of SE for the case of vanishing magnetic field. Furthermore, the solutions with separated variables are of the form (6) with $Q = \exp(iS)$, where $S = S(t,\vec{x})$ is given by (49), the functions $\omega_1(t,\vec{x}), \omega_2(t,\vec{x}), \omega_3(t,\vec{x})$ are given implicitly by formulas (48), (22), and (4). Again, we have the eleven classes of vector potentials $A(t,\vec{x})$ corresponding to the eleven classes of coordinate systems $\omega_a = \omega_a(t,\vec{x})$, $a = 1,2,3$. SE (1) for each class of the functions $A_0(t,\vec{x}), \vec{A}(t,\vec{x}) = \vec{0}$ defined by (50) under arbitrary $\tilde{T}_0(t), F_{a0}(\omega_a)$ and fixed arbitrary functions $w_a(t), h_a(t)$, $a = 1,2,3$ separates in exactly one coordinate system. The separation equations read as (11), where the coefficients $F_{ai}$, $a,i = 1,2,3$ are the entries of the corresponding Stäckel matrices (30), functions $T_a$, $a = 1,2,3$ are listed in (34) and the functions $T_0, F_{a0}, a = 1,2,3$ are arbitrary smooth functions defining the form of the timelike component of the vector potential $A(t,\vec{x})$ (see, (50)).

If we fix the temporal variable $t$ to be equal to $t_0 \in \mathbf{R}$ in the above obtained results, then it is not difficult to classify separable stationary Schrödinger equations for a particle interacting with the electromagnetic field $A(\vec{x}) = (A_0(\vec{x}), \vec{A}(\vec{x}))$,

$$(p_a p_a + eA_0 + E)\psi(\vec{x}) = 0, \tag{51}$$

where $p_a = i\partial/\partial x_a - eA_a$, $a = 1, 2, 3$, $e = \mathrm{const}$, and $E$ is a spectral parameter, and thus recover the classical result by Eisenhart [9].

## III. Separation of variables in the Hamilton–Jacobi equation

It is well known that there exists a deep connection between the separation of variables in the Schrödinger and Hamilton–Jacobi equations (see, e.g., [16]). The Hamilton–Jacobi equation,

$$u_t + eA_0 + (u_{x_a} + eA_a)(u_{x_a} + eA_a) = 0, \tag{52}$$

separates in any coordinate system providing separability of the Schrödinger equations (1) and, what is more, the inverse assertion is not true. We will make use of this connection for the sake of classifying separable Hamilton–Jacobi equations.

First we fix the usual form of the separation Ansatz for the Hamilton–Jacobi equation,

$$u(t, \vec{x}) = S(t, \vec{x}) + \varphi_0(t) + \sum_{i=1}^{3} \varphi_i(\omega_i(t, \vec{x})), \tag{53}$$

and, furthermore, fix the form of the ordinary differential equations for $\varphi_0, \varphi_1, \varphi_2, \varphi_3$,

$$\varphi_0' = -T_0(t) - T_i(t)\lambda_i, \quad \varphi_a' = \left(-F_{a0}(\omega_a) + F_{ai}(\omega_a)\lambda_i\right)^{1/2}. \tag{54}$$

Now, inserting the Ansatz (53) into equation (52), eliminating the first derivatives of the functions $\varphi_0, \varphi_1, \varphi_2, \varphi_3$ with the use of the above equations and splitting by the variables $\varphi_0, \varphi_1, \varphi_2, \varphi_3, \lambda_1, \lambda_2, \lambda_3$, we arrive at the following system of nonlinear partial differential equations for the functions $S, \omega_1, \omega_2, \omega_3$:

$$\frac{\partial \omega_i}{\partial x_a} \frac{\partial \omega_j}{\partial x_a} = 0, \quad i \neq j, \quad i,j = 1,2,3;$$

$$\sum_{i=1}^{3} F_{ia}(\omega_i) \frac{\partial \omega_i}{\partial x_j} \frac{\partial \omega_i}{\partial x_j} = T_a(t), \quad a = 1,2,3;$$

$$2\left(\frac{\partial S}{\partial x_j} + eA_j\right) \frac{\partial \omega_a}{\partial x_j} + \frac{\partial \omega_a}{\partial t} = 0, \quad a = 1,2,3; \qquad (55)$$

$$-\sum_{i=1}^{3} F_{i0}(\omega_i) \frac{\partial \omega_i}{\partial x_j} \frac{\partial \omega_i}{\partial x_j} + \frac{\partial S}{\partial t} + 2eA_a \frac{\partial S}{\partial x_a} + \frac{\partial S}{\partial x_a} \frac{\partial S}{\partial x_a}$$
$$-T_0(t) + eA_0 + e^2 A_a A_a = 0.$$

The general solution $\vec{\omega} = \vec{\omega}(t, \vec{x})$ of the first six equations of the above system (which coincide with equations (12) and (13)) can be reduced with the help of an appropriate equivalence transformation $\mathcal{E}$ to such a form that it satisfy the Laplace equation (5).

It is not difficult to become convinced of the fact that making the change of variables,

$$Q(t, \vec{x}) = \exp(iS(t, \vec{x})), \qquad (56)$$

in (12)–(15) yields the system that coincides with (55) with an exception of the last equation, where an additional term $-i(\Delta S + eA_{ax_a})$ appears. As shown in the previous section, this term is a function of $t$ only and is absorbed by $T_0$. Consequently, all the results on variable separation for SE apply to the case of the Hamilton–Jacobi equation (52) as well.

## IV. Concluding Remarks

Theorems 1–2 give a complete solution of the problem of classification of SE's (1) and (51) that are solvable within the framework of the method of separation of variables. By appropriate reductions of these results we can get the results on the separation of variables in SE's for a particle interacting with the electromagnetic field in one [24] and two [26, 27] spatial dimensions. For example, in order to recover the results of [27] one has to choose $\partial \psi / \partial x_3 = 0$ and consider the completely and partially split coordinate systems from the list (4).

It follows from Theorem 1 that the choice of magnetic fields $\vec{H}$ allowing for variable separation in the corresponding SE is very restricted. Namely, the magnetic field should be independent of spatial variables $x_1, x_2, x_3$ in order to provide the separability of SE (1) into three second-order ordinary differential equations. However, if we allow for separation equations to be of lower order, then additional possibilities for variable separation in SE arise. As an example we give the vector potential,

$$A(t, \vec{x}) = \left(A_0\left(\sqrt{x_1^2 + x_2^2}\right), 0, 0, A_3\left(\sqrt{x_1^2 + x_2^2}\right)\right),$$

where $A_0, A_3$ are arbitrary smooth functions. SE (1) with this vector-potential separates in the cylindrical coordinate system $t, \omega_1 = \ln\left(\sqrt{x_1^2 + x_2^2}\right), \omega_2 = \arctan(x_1/x_2), \omega_3 = x_3$ into two first-order and one second-order ordinary differential equations. The corresponding magnetic field $\vec{H} = \operatorname{rot} \vec{A}$ is evidently $x$ dependent.

As mentioned in the Introduction, a possibility of variable separation in SE is intimately connected to its symmetry properties. Namely, solutions with separated variables are common eigenfunctions of three mutually commuting symmetry operators of SE. For all the cases of variable separation in SE (1) these operators can be constructed in explicit form, in analogy to what has been done in [26] for the (1+2)-dimensional case. They are expressed in terms of the coefficients of the separation equations (11). This fact enables application of the methods of the representation theory of Lie algebras for a further analysis of special functions arising as solutions of separation equations in the spirit of the famous Bateman's project [15].

The last remark is that the technique developed in the present paper can be directly applied in order to separate variables in the Pauli equation for a particle with spin $\frac{1}{2}$ moving in the electromagnetic field and in the Fokker–Planck equation with a constant diagonal diffusion matrix.

A study of the above-mentioned problems is in progress now and will be reported in our future publications.